\documentstyle[seceq,preprint]{jpsj}

\newcommand{\Rvec}{\mbox{\boldmath $R$}}
\newcommand{\gvec}{\mbox{\boldmath $g$}}

\newcommand{\pvec}{\mbox{\boldmath $p$}}
\newcommand{\rvec}{\mbox{\boldmath $r$}}
\newcommand{\uvec}{\mbox{\boldmath $u$}}
\newcommand{\vvec}{\mbox{\boldmath $v$}}
\newcommand{\qvec}{\mbox{\boldmath $q$}}
\newcommand{\svec}{\mbox{\boldmath $s$}}
\newcommand{\yvec}{\mbox{\boldmath $y$}}
\newcommand{\zvec}{\mbox{\boldmath $z$}}
\newcommand{\wfvec}{\mbox{\boldmath $\Psi$}}
\newcommand{\NB}{N_{\mbox{\tiny B}}}
\newcommand{\NG}{N_{\mbox{\tiny G}}}
\newcommand{\NENE}{N_{\mbox{\tiny E}}}
\newcommand{\NBIT}{N_{\mbox{\tiny bit}}}
\newcommand{\NITER}{N_{\mbox{\tiny iter}}}
\newcommand{\IMAX}{I_{\mbox{\tiny max}}}
\newcommand{\PINT}{\pvec_{\mbox{\tiny I}}}

\begin{document}

\title{An efficient algorithm for electronic-structure calculations}

\author{E\hspace{0.4mm}i\hspace{0.4mm}j\hspace{0.4mm}i Tsuchida}

\inst{Research Institute for Computational Sciences, AIST, \\
Tsukuba Central 2, Umezono 1-1-1, Tsukuba, Ibaraki 305-8568, Japan}

\abst{
We show how to adapt the quasi-Newton method
to the electronic-structure calculations
using systematic basis sets.
Our implementation requires less iterations than
the conjugate gradient method, while the computational
cost per iteration is much lower.
The memory usage is also quite modest,
thanks to the efficient representation of the
approximate Hessian.
}

\kword{density-functional theory, quasi-Newton method,
BFGS update, finite-element method,
Born-Oppenheimer dynamics}

\maketitle

\section{Introduction}

The importance of the first-principles
electronic-structure calculations
based on the density-functional theory \cite{HK,KS,CP}
is increasing year by year \cite{PAY,TREV,MARX}.
Since the optimization of the ground-state wavefunctions is
the most time-consuming part of these calculations,
it is crucial to use an efficient algorithm for this purpose.
However, the number of degrees of freedom
is so large for systematic basis sets like
plane-waves \cite{PAY,TREV,MARX},
finite-differences \cite{CTS,BER,RREV},
and finite-elements \cite{RREV,WHT,PRB2,JPSJ,PKFS}, that
the memory usage of the algorithm being used
is severely restricted.
Currently, the conjugate gradient method \cite{RCP,GLL,TPA,SCP,BKL,PAY}
seems to be most widely used because of
its efficiency and modest memory usage,
while the direct inversion in the iterative subspace
(DIIS) \cite{PLY,WOZU,MACO,HLP,KRFU} is also sometimes used.

On the other hand, the quasi-Newton methods have rarely been used
for electronic optimization
in combination with systematic basis sets, although their efficiency
is well known \cite{SREV,FREV,NREV};
to the best of our knowledge, the application of the quasi-Newton
methods in this context
has been limited to atomic orbitals \cite{HEPO,FIAL} or
one-dimensional problems \cite{HSZ}.
This is presumably because they require significantly more storage
for the elements of the (approximate) Hessian matrix.
If an all-band update is used,
the dimension of the Hessian (${\cal H}$) is given by
${\cal N} = \NB \NG$, where $\NB$ is the number of
orbitals and $\NG$ is the number of basis functions.
Therefore, the storage requirement for
${\cal H} \, ({\cal N} \times {\cal N}) $ will be
${\cal N}^2$ in a naive implementation \cite{RCP}, which is
prohibitive for large-scale simulations
where $\cal N$ can exceed 10$^7$.
A more practical implementation of the quasi-Newton method is
also found in the literature \cite{NOCE}, in which
only the $m$ previous steps are relevant.
Since two update vectors of size $\cal N$ are required
per step \cite{NOCE},
the memory usage amounts to $2 m {\cal N}$ elements,
where $m$ is usually less than 10.
However, this can be further reduced to $ m {\cal N}$
if the initial Hessian
is a multiple of the unit matrix \cite{SIEG,GILE}. 
In this article, we present the implementation of
the quasi-Newton method using
the BFGS (Broyden-Fletcher-Goldfarb-Shanno) formula \cite{FREV,GILE}
along this line.
As explained in the next section,
we make a number of modifications to adapt the algorithm
to the electronic-structure calculations.
The most important one is the compression of the
update vectors by an order of magnitude,
which makes this algorithm attractive
even for very large systems.

\section{Methods}

\subsection{Electronic-structure calculations}
First of all, we explain the basic problems
in the electronic-structure
calculations within the density-functional theory \cite{HK,KS}.
Only real wavefunctions at the $\Gamma$-point of the Brillouin zone
are considered for notational simplicity,
but generalization to complex wavefunctions is straightforward.

The total energy functional for an ionic configuration $\Rvec$
is given by \cite{PAY}
\begin{eqnarray}
E_{\mbox{\scriptsize total}} \, [\wfvec, \Rvec] & = & 
\sum_i \int \psi_i (\rvec) \left[ -\nabla^2 +
V_{\mbox{\scriptsize ps}} [\Rvec] \right] \psi_i (\rvec) \, {\mbox d}\rvec
+ E_{\mbox{\scriptsize Hxc}} [n (\rvec)]
+ E_{\mbox{\scriptsize ion}} [\Rvec],
\end{eqnarray}
where 
\begin{equation}
\wfvec=(\psi_1(\rvec) \;\; \psi_2(\rvec) \;\;... \;\; \psi_{\NB}(\rvec))^T, 
\end{equation}
\begin{equation}
n (\rvec) = \sum_i |\psi_i (\rvec)|^2,
\end{equation}
and $E_{\mbox{\scriptsize Hxc}}$ is the sum of
the Hartree and exchange-correlation
energy, which is a nonlinear and nonlocal functional of
the electron density $ n (\rvec) $.
In practice, each $\psi_i(\rvec)$ is discretized by a
basis set expansion \cite{PAY,TREV,MARX,CTS,BER,WHT,RREV,PRB2,JPSJ,PKFS},
which makes $\wfvec$ a huge vector with ${\cal N} (=\NB \NG)$ elements. 

In the conventional approach \cite{PAY},
the ground-state energy $E_{\mbox{\tiny G}}$ and
wavefunctions $\wfvec_{\mbox{\tiny G}}$ for the given $\Rvec$
are obtained by minimization of
$ E_{\mbox{\scriptsize total}}[\wfvec, \Rvec] $
with respect to the wavefunctions $\wfvec$ 
under the orthonormality constraints: 
\begin{equation}
\int \psi_i (\rvec) \, \psi_j (\rvec) \, {\mbox d}\rvec = \delta_{ij}. 
\end{equation}
$\wfvec_{\mbox{\tiny G}}$ calculated in this way is then used to study
various properties of the system. 

In our implementation, on the other hand,
the above constraints are eliminated
by modifying the total energy functional
according to Refs. [\citen{SCP,APJ1,MGC,KV}],
in which orthonormality of the wavefunctions is satisfied
either implicitly \cite{SCP,APJ1,KV} or automatically \cite{MGC}.
Moreover, all the orbitals are updated
simultaneously \cite{KV,JPSJ}, and
self-consistency of $E_{\mbox{\scriptsize Hxc}}$ is taken into account
in the evaluation of its gradient. 
Then, if the modified total energy functional for the given $\Rvec$
is denoted by $E \, [\wfvec]$, $E_{\mbox{\tiny G}}$
and $\wfvec_{\mbox{\tiny G}}$ are obtained by
minimization of $E \, [\wfvec]$ with respect to $\wfvec$
without any constraints. 
Thanks to this reformulation, we can easily implement
the quasi-Newton method which is one of the most efficient algorithms
for the unconstrained optimization of
nonlinear functions \cite{SREV,FREV,NREV}.
Furthermore, the use of nonorthogonal basis functions is
much easier in this case \cite{JPSJ,KV}.
The above ground-state calculations are usually performed
for a series of slowly varying $\Rvec$, each of which
is called an {\it ionic step}.

\subsection{BFGS with full Hessian}
\label{FBFGS}
We illustrate the conventional 
quasi-Newton method using the BFGS formula \cite{FREV,GILE} here,
which will serve as a prototype
for the implementation in reduced space.
For simplicity, we assume the new total energy
(eq. (\ref{ENEW})) is always lower than the previous value. \\
\ \\
Choose ${\cal H}_0$ and $\wfvec_0$. \\
Calculate $ E_0 = E \, [\wfvec_0]$ and
$\gvec_0 = \nabla E \, [\wfvec_0]$. \\
Set k=0. \\
Do while ($|\gvec_k| \ge \epsilon$)
\begin{equation}
\pvec_k = -{\cal H}_k^{-1} \gvec_k
\end{equation}
\begin{equation}
\wfvec_{k+1} = \wfvec_k + \pvec_k
\end{equation}
\begin{equation}
\label{ENEW}
E_{k+1} = E \, [\wfvec_{k+1}]
\end{equation}
\begin{equation}
\gvec_{k+1} = \nabla E \, [\wfvec_{k+1}]
\end{equation}
\begin{equation}
\Delta \wfvec_k = \wfvec_{k+1} - \wfvec_k
\end{equation}
\begin{equation}
\Delta \gvec_k = \gvec_{k+1} - \gvec_k
\end{equation}
\begin{equation}
\label{FBFGSEQ}
{\cal H}_{k+1} = {\cal H}_k -
\frac{{\cal H}_k \Delta \wfvec_k \Delta \wfvec_k^T {\cal H}_k}
{\Delta \wfvec_k^T {\cal H}_k \Delta \wfvec_k}
+\frac{\Delta \gvec_k \Delta \gvec_k^T}
{\Delta \wfvec_k^T \Delta \gvec_k}
\end{equation}
\begin{equation}
k=k+1
\end{equation}
End do \\

While this algorithm is simple and efficient in terms of the
convergence rate, its memory usage and computational effort
scale as $O({\cal N}^2)$ and $O({\cal N}^3)$ respectively,
which are prohibitive.
Although the latter can be reduced to $O({\cal N}^2)$, if the
updating formula for the inverse Hessian (${\cal H}^{-1}$)
is used \cite{RCP}, this is still far from practical.
The purpose of this article is to present
the improved algorithm \cite{SIEG,GILE}
in which both scale as $O({\cal N})$ with modest prefactors.

\subsection {QR-decomposition}
\label{QRD}
At this point, we give a brief introduction to the
QR-decomposition \cite{RCP},
which plays an important role in the algorithm
presented in the next section.
Let us assume $B ({\cal N} \times r)$ is a set of
linearly independent vectors:
\begin{equation}
B = (\pvec_1 \;\; \pvec_2 \;\; \cdots \;\; \pvec_r), 
\end{equation}
where $1 \le r \ll {\cal N}$.
Then the QR-decomposition of $B$ is given by 
\begin{equation}
B = Z \,\, T, 
\end{equation}
where
$Z ({\cal N} \times r)$ is a set of orthonormal vectors spanning the
same subspace as $B$, i.e. 
\begin{equation}
Z^T Z = I, 
\end{equation}
and $T (r \times r)$ is an invertible upper-triangular matrix.
In practice, this decomposition is obtained by applying the
addition procedure given below repeatedly, which is
(mathematically) equivalent to constructing an orthonormal basis
from the left ($\pvec_1$) to the right ($\pvec_r$)
by the Gram-Schmidt scheme.
Note, however, that only $B$ and $T$ are considered explicitly
in the following \cite{SIEG,GILE}.

Here we show how to update the above QR-decomposition
when $B$ is slightly modified.
In the first case where a vector $\gvec$ is added to $B$, i.e.
\begin{equation}
B_+ = (\pvec_1 \;\; \pvec_2 \;\; \cdots \;\; \pvec_r \;\; \gvec)
= (B \;\; \gvec), 
\end{equation}
the new decomposition is given by 
\begin{equation}
B_+ = Z_+ \, T_+, 
\end{equation}
where 
\begin{equation}
T_+ ((r+1) \times (r+1)) = \left(
\begin{array}{cc}
T &  \uvec \\
0 &  \rho \\
\end{array}
\right) ,
\end{equation}
\begin{equation}
\uvec = Z^T \gvec = (T^T)^{-1} ( B^T \gvec ), 
\end{equation}
and 
\begin{equation}
\rho = \sqrt{|\gvec|^2 - |\uvec|^2}.
\end{equation}
If $\rho \ne 0$, $T_+$ is also an invertible upper-triangular
matrix.

Next, we consider the case of
dropping the leftmost vector $\pvec_1$ from $B$, i.e.
\begin{equation}
B_- = (\pvec_2 \;\; \pvec_3 \;\; \cdots \;\; \pvec_r).
\end{equation}
The corresponding decomposition is given by 
\begin{equation}
B_- = Z_- T_-,
\end{equation}
where $T_-$ satisfies
\begin{equation}
\label{QRM}
T^T_- \, T_- = B^T_- \, B_-.
\end{equation}
Obviously, the right-hand side of eq. (\ref{QRM}) is
included in $ B^T B$, which is easily calculated from  
\begin{equation}
B^T B = T^T T.
\end{equation}
Therefore, $T_-$ is obtained by the Cholesky decomposition \cite{RCP}
of a small matrix at negligible cost.
A more refined approach is introduced
in Ref. \citen{GILE}, but
the above procedure seems to be sufficient for our present purpose.

\subsection{BFGS with reduced Hessian and limited memory}
\label{ALG}
Here we present the state-of-the-art implementation of the
quasi-Newton method \cite{SIEG,GILE}, which is obtained by
modifying the conventional algorithm ($\S$ \ref{FBFGS})
under two assumptions: (i) ${\cal H}_0 = \sigma I$ ($\sigma > 0$), and
(ii) At most $m$ previous steps are stored.

In order to fully exploit these conditions, it is more
convenient to use a compact representation for the
Hessian: 
\begin{equation}
H = Z^T \, {\cal H} \, Z,
\end{equation}
where $Z \, ({\cal N} \times r)$ is the current (orthonormal) basis,
$H \, (r \times r)$ is the reduced Hessian,
and $1 \le r \le m+1 \ll {\cal N}$.
While $Z$ and ${\cal H}$ also appear in the following algorithm,
they are not explicitly calculated.
The reduced vectors are defined in a similar way;
the reduced gradient $\uvec$, for instance, is given by
$\uvec = Z^T \gvec$, where $\gvec = \nabla E$. 
The correspondence between the full/reduced vectors
is shown in Table \ref{TAB0}. 

\begin{enumerate}
\item \label{QNINI} Initilization: \\
Set $k=0$ and $r=1$, where $k$ and $r$ denote the loop index
and the rank of the reduced space, respectively. \\
Choose the initial wavefunction ($\wfvec_0$),
the approximate curvature ($\sigma$),
the convergence criterion ($\epsilon$), and
the maximum rank of the reduced space ($m$). \\
Calculate the total energy
\begin{equation}
E_0=E \, [\wfvec_0]
\end{equation}
and its gradient 
\begin{equation}
\gvec_0=\nabla E \, [\wfvec_0].
\end{equation}
{\sf IF} ($|\gvec_0| < \epsilon$) {\sf THEN} quit. \\
{\sf ELSE} $ H_0 = (\sigma), \; B_0=(\gvec_0), \;
T_0=(|\gvec_0|)$, and $ \vvec_0=(|\gvec_0|) $. Moreover,
$Z_0 = (\gvec_0/|\gvec_0|)$ and ${\cal H}_0 = \sigma I$
are implicitly assumed. \\
If the Hessian of the previous ionic step is
taken over, several modifications are
required in this step, which are, however, straightforward.

\item \label{QNLOOP} Calculate the new search direction in reduced space:
\begin{equation}
\label{QKEQ}
\qvec_k = -H_k^{-1} \vvec_k.
\end{equation}
\item \label{PFUL} Calculate the new search direction:
\begin{equation}
\label{FULLP}
\pvec_k = Z_k \qvec_k = B_k (T_k^{-1} \qvec_k).
\end{equation}
\item \label{UPB1} Update the subspace: \
$ (B_k = Z_k T_k \rightarrow B_k' = Z_k T_k') $,
where
\begin{equation}
B_k ({\cal N} \times r) =
(\pvec_{k-r+1} \;\; \cdots \;\; \pvec_{k-1} \;\; \gvec_k)
\end{equation}
and
\begin{equation}
B'_k ({\cal N} \times r) =
(\pvec_{k-r+1} \;\; \cdots \;\; \pvec_{k-1} \;\; \pvec_k).
\end{equation}
$T_k'$ is obtained from $T_k$ and $\qvec_k$,
whereas $Z_k$ remains unchanged \cite{GILE}.
\item Set $\alpha = 1$ and calculate the gradient of the total energy
along $\pvec_k$ as
\begin{equation}
\label{EDEQ}
E' = \left. 
\frac{\partial E \, [\wfvec_k + \alpha \pvec_k] }{\partial \alpha}
\right|_{\alpha=0}
= \gvec_k^T \pvec_k = \vvec_k^T \qvec_k.
\end{equation}
\item \label{QNLINMIN} Calculate the new wavefunction: 
\begin{equation}
\wfvec_{k+1} = \wfvec_k + \alpha \pvec_k.
\end{equation}
\item \label{QNLM2} Calculate the new total energy: 
\begin{equation}
E_{k+1} = E \, [\wfvec_{k+1}].
\end{equation}
{\sf IF} $(E_{k+1} \geq E_k)$ {\sf THEN}
estimate the optimal $\alpha$ by a parabolic fit
with $E_k$, $E'$, and $E_{k+1}$, and go to \ref{QNLINMIN}.
\item \label{GFUL} Calculate the new gradient:
\begin{equation}
\gvec_{k+1}=\nabla E \, [\wfvec_{k+1}].
\end{equation}
{\sf IF} ($|\gvec_{k+1}| < \epsilon$) {\sf THEN} quit.

\item \label{RGITEM} Extend the subspace: \
$( B_k'=Z_k T_k' \rightarrow B_k''= Z_k' T_k'') $,
where 
\begin{equation}
B'_k({\cal N} \times r) =
(\pvec_{k-r+1} \;\; \cdots \;\; \pvec_k)
\end{equation}
and
\begin{equation}
B_k'' ({\cal N} \times (r+1)) =
(\pvec_{k-r+1} \;\; \cdots \;\; \pvec_{k} \;\; \gvec_{k+1}).
\end{equation}
As explained in $\S$ \ref{QRD},
$T_k''$ is obtained from $T_k'$, $\uvec_k$, and $\rho_{k+1}$, where
\begin{equation}
\label{REDG}
\uvec_{k} = Z_k^T \gvec_{k+1}=(T_{k}^{'T})^{-1} (B_{k}^{'T} \gvec_{k+1})
\end{equation}
and 
\begin{equation}
\rho_{k+1} = \sqrt{|\gvec_{k+1}|^2 - |\uvec_k|^2}.
\end{equation}
We assume $\rho_{k+1} \ne 0$ in the following. 
Then, the new basis $Z_k' ({\cal N} \times (r+1))$
is given by \cite{GILE}
\begin{equation}
Z_k' = ( Z_k \;\;\; \zvec_{k+1}), 
\end{equation}
where $\zvec_{k+1} = (\gvec_{k+1} - Z_k \uvec_k) / \rho_{k+1}$.
However, $\zvec_{k+1}$ is not explicitly calculated. 

\item $ r=r+1 $
\item Calculate the reduced gradients as 
\begin{equation}
\vvec_k' = Z_k^{'T} \gvec_k = \left(
\begin{array}{c}
\vvec_k \\
0 \\
\end{array} \right)
\end{equation}
and 
\begin{equation}
\uvec_k' = Z_k^{'T} \gvec_{k+1} = \left(
\begin{array}{c}
\uvec_k \\
\rho_{k+1}
\end{array}
\right). 
\end{equation}
There is no loss of information here, since
$\gvec_k, \gvec_{k+1} \in Z_k'$. 

\item \label{QNSY}
Update the reduced Hessian using the BFGS formula \cite{FREV,GILE}:
\begin{equation}
\label{RBFGSEQ}
H''_k (r \times r)
= Z_k^{'T} {\cal H}_{k}^{+} Z'_k
= H'_k
- \frac{H'_k \svec_k \svec_k^T H'_k}{\svec_k^T H'_k \svec_k}
+ \frac{\yvec_k \yvec_k^T}{\svec_k^T \yvec_k},
\end{equation}
where 
\begin{equation}
\svec_k = Z_{k}^{'T} \Delta \wfvec_k =
\alpha \left(
\begin{array}{c}
\qvec_k \\
0 \\
\end{array}
\right),
\end{equation}
\begin{equation}
\yvec_k = Z_{k}^{'T} \Delta \gvec_k = \uvec_k' - \vvec_k',
\end{equation}
and
\begin{equation}
\label{SIG2}
H'_{k}(r \times r) = Z_k^{'T} {\cal H}_{k} Z'_k = 
\left(
\begin{array}{cc}
H_{k}  & 0 \\
0    & \sigma \\
\end{array}
\right).
\end{equation}
${\cal H}_k^+$ is defined as the right-hand side of
eq. (\ref{FBFGSEQ}), and eq. (\ref{RBFGSEQ}) is derived from
this definition.
Note that $\svec_k^T \yvec_k$ $ > 0$ is assumed here; otherwise,
the Hessian is not updated.

\item {\sf IF} $(r = m+1)$ {\sf THEN}
reduce the subspace: \ $(B_k'' = Z_k' T_k''
\rightarrow B_{k+1} = Z_{k+1} T_{k+1})$, where
\begin{equation}
B_k'' ({\cal N} \times (m+1)) =
(\pvec_{k-m+1} \;\; \pvec_{k-m+2} \;\;
\cdots \;\; \pvec_{k} \;\; \gvec_{k+1})
\end{equation}
and
\begin{equation}
B_{k+1} ({\cal N} \times m) =
(\pvec_{k-m+2} \;\; \cdots \;\; \pvec_{k} \;\; \gvec_{k+1}).
\end{equation}
$T_{k+1}$ is easily obtained from $T_k''$ according to $\S$ \ref{QRD}. 
Then, $H_{k+1} (m \times m) = Z_{k+1}^T {\cal H}_k^+ Z_{k+1}$ 
is calculated from $T_k'', T_{k+1}$, and $H_k''$
by way of $B_{k+1}^T {\cal H}_k^+ B_{k+1}$.
At this point, the new Hessian (${\cal H}_{k+1}$)
in the new basis ($Z_{k+1}$ and its orthogonal complement)
is defined as a block-diagonal matrix consisting of $H_{k+1}$
and $\sigma I (({\cal N}-m) \times ({\cal N}-m))$,
which was implicitly used in eq. (\ref{SIG2}).
Therefore, part of the information contained in ${\cal H}_k^+$
has been discarded here.
Similarly, $\vvec_{k+1} = Z_{k+1}^T \gvec_{k+1}$ is calculated
from $T_k'', T_{k+1}$, and $\uvec_k'$
by way of $B_{k+1}^T \gvec_{k+1}$, but there is no loss of
information here.
Finally, we set $r=m$. \\
{\sf ELSE}
$ H_{k+1}=H_k'', B_{k+1}=B_k'', T_{k+1}=T_k''$,
and $\vvec_{k+1}= \uvec_k' $.
Moreover, $Z_{k+1} = Z_k'$ and ${\cal H}_{k+1} = {\cal H}_k^+$
are implicitly assumed.
\item $k=k+1$
\item Go to \ref{QNLOOP}.
\end{enumerate}

\begin{itemize}
\item While $0 \le k \le m-1$,
this algorithm is identical to the conventional one
($\S$ \ref{FBFGS}) with ${\cal H}_0 = \sigma I $
within round-off errors.
The two algorithms begin to differ once $k$ reaches $m$,
but the deterioration of the convergence rate is minimized
by constructing the subspace with
the previous search directions rather than the
gradients \cite{SIEG,GILE}.

\item For simplicity,
the above algorithm includes minimal exception handling.
Therefore, the original paper \cite{GILE} should be
consulted for a more complete one. 
However, such exceptions are observed
only in the very early stages of the first
ionic step, where the quadratic model is not valid.

\item One cycle requires approximately
$ 2 r {\cal N}$ multiply-and-add operations,
arising from eqs. (\ref{FULLP}) and (\ref{REDG}).
For practical values of $m \, (< 10$), these costs will be
much lower than those of evaluating
the total energy in step \ref{QNLM2} \cite{PAY}.

\item The basis functions should be appropriately
scaled \cite{TPA,HEPO} in advance,
so that their contribution to the total energy is similar.

\item The reduced Hessian $H_k$ is diagonalized in each cycle to
guarantee its positive definiteness; nonpositive eigenvalues,
if any, are modified appropriately.
Then, it follows from eqs. (\ref{QKEQ}) and (\ref{EDEQ}) that
$ E' = - \vvec_k^T H_k^{-1} \vvec_k < 0$,
because $H_k^{-1}$ is also positive definite and
$|\vvec_k| = |\gvec_k| \ge \epsilon$.
Furthermore,
the average eigenvalue of the reduced Hessian,
denoted by $\lambda_k$, is also calculated and stored for later use.

\item We explain the choice of $\sigma$
used in step \ref{QNINI} and \ref{QNSY} here.
Since $\sigma$ is the approximate curvature
along the new direction \cite{GILE}, a reasonable
estimate is needed to achieve high performance. 
Therefore, a number of strategies have been proposed 
to choose optimal $\sigma$ \cite{NOCE,SIEG,GILE},
most of which provide dynamical estimates.
Nevertheless, we use a constant $\sigma$ during each ionic step
unless otherwise noted, which is determined as follows:
In the first ionic step, $\sigma$ is estimated from the
coarse grid iterations \cite{JPSJ}.
At the end of each ionic step, the sequence $\{ \lambda_k \}$
is further averaged
to give the new $\sigma$ for the next ionic step.
$\sigma$ obtained in this way varies only slowly with ionic steps,
while providing stable and high performance
in the systems we have studied so far.
Comparison is also made with the dynamical estimates
in $\S$ \ref{RESSEC}.
\end{itemize}

\subsection{Data compression}
\label{ZIP}
The memory usage of the algorithm illustrated in the
previous section is dominated by the $m$ previous search directions,
which amount to $m {\cal N}$ elements.
While this is much smaller than
the storage of the full Hessian $(={\cal N}^2)$, it is still
a serious obstacle in large-scale simulations.
In what follows, we present a simple algorithm to compress
the previous search directions
without sacrificing the efficiency of the original method.
In this algorithm, one search direction is compressed
in each cycle, by taking advantage of its structure.
If $ \pvec \, ({\cal N}) $, which is being compressed,
is viewed as a two-dimensional array $ \pvec \, (\NB, \NG) $,
the size of $\pvec \, (i, j)$
for a given basis function ($j$) is
expected to be similar for all orbitals ($i$).
Based on this idea, the largest element of $ |\pvec (i, j)|$
with respect to $i$ is chosen as the scale factor. 
Moreover, $\NBIT$ is defined as the number of bits
assigned to each element of $\pvec$ after compression.  

Then, the scale factor $\omega \, (\NG)$
and the compressed array $\PINT \, (\NB, \NG)$ are given by
\begin{equation}
\label{SCLF}
\mbox{real$\ast$8} \;\;\;\;\; \omega \, (j) =
\left( \max_{1 \le i \le \NB} |\pvec \, (i,j)| \right) / \IMAX
\end{equation}
and 
\begin{equation}
\mbox{integer}  \;\;\;\;\; \PINT \, (i,j)=
\mbox{round} \left( \frac{\pvec \, (i,j)}{\omega \, (j)} \right)
+ \IMAX
\end{equation}
respectively,
where $ \IMAX = 2^{\NBIT-1} -1 $
and $ 0 \le \PINT \, (i,j) \le 2 \IMAX = 2^{\NBIT} -2 $.
Therefore, each element of $\PINT$ is representable by
$\NBIT$ bits. 
The original values of $\pvec$ are recovered approximately by 
\begin{equation}
\pvec \, (i,j) \approx \omega (j) \, (\PINT (i,j) - \IMAX).
\end{equation}

In this method, the quality of the compression can
be controlled by a single parameter, $\NBIT$.
Furthermore, the largest element for each $j$,
which is the most important one, remains exact. 

The total storage for the $m$ search directions 
after compression is $ m \NBIT {\cal N} / 8 $ bytes,
if appropriately packed with bit operations.
If $m=\NBIT=8$, for instance, this amounts to only
one double-precision array of size $\cal N$.
Note also that the storage for the scale factors is minor.

In the current implementation,
$\pvec_k$ is compressed in step \ref{UPB1}, when added to $B_k'$. 
At the same time, the last column of $T_k'$ is
calculated directly from the compressed $\pvec_k$
(rather than using $\qvec_k$)
to maintain the consistency of the QR-decomposition.
However, the uncompressed $\pvec_k$ is also retained and used
in step \ref{QNLINMIN}.

Unfortunately, some inconsistency seems inevitable
in the update of the reduced Hessian, since
$\svec_k$ and $\yvec_k$ no longer belong to $Z_k'$.
Nevertheless,
$E' < 0$ remains valid as long as the reduced Hessian is positive
definite and the latest $\gvec$ and $\pvec$ are uncompressed.
Therefore, the stability of the minimization is guaranteed
even if the previous search directions are highly compressed.

\section{Results}
\label{RESSEC}
As a test of our implementation under realistic conditions,
we performed a series of Born-Oppenheimer dynamics \cite{WM} 
for bulk diamond at 220 K in a periodic cubic supercell of
64 atoms within the local density approximation \cite{HK,KS}.
The wavefunctions were expanded by the adaptive finite-element
method \cite{PRB2,JPSJ} with an average cutoff energy of 43 Ry,
which corresponds to $ \NG = 8 \times 14^3 = 21,952$.
Since $\NB$ is equal to 128,
${\cal N}$ amounts to approximately 2,800,000 in this system.
The Brillouin zone was sampled only at the $\Gamma$-point,
and the separable pseudopotentials were used \cite{KB,GTH}.
The convergence criterion ($\epsilon$) was chosen so that
$|E_{k+1} - E_k| \simeq 2 \times 10^{-8}$ Ry/atom
when $|\gvec_{k+1}| < \epsilon$ was satisfied. 
Convergence to the ground state was accelerated by the
enhanced extrapolation scheme \cite{EES},
which provides accurate
initial wavefunctions with the help of population analysis.

The equations of motion for the ions were integrated using
the velocity-Verlet method \cite{LIQ}
with a timestep of 80 a.u. ($\sim 2$ fs). 
Starting from the same ionic configuration,
each run lasted for 57 ionic steps,
the last 50 steps of which were used to collect the statistics.
Moreover, $B,T$, and $H$ were taken over from
previous ionic steps unless otherwise noted.
Therefore, these matrices were saturated during this period in all runs.

We first show the average number of iterations
($\NITER$) and total energy evaluations ($\NENE$)
needed to optimize the electronic-structures
for the conjugate gradient method
using the Polak-Ribiere formula \cite{RCP} and
the quasi-Newton method using the BFGS formula
in Table \ref{TAB1}.
The convergence rate of the quasi-Newton method
as measured by $\NITER$ is already comparable to that of
the conjugate gradient method for $m = 2$,
and becomes better as $m$ is increased.
However, there is no point in using $m$ much larger
than $\NITER$ (say, 20), because the Hessian
is dominated by the contribution from previous ionic steps.
In practice,
any reasonable choice of $m$, e.g. 5-8, will provide near-optimal
performance, since $\NITER$ depends
only weakly on $m$ in this range.
Note also that the CPU-time is more closely related to
$\NENE$ than $\NITER$.
Therefore, the quasi-Newton method was much faster
than the conjugate gradient method for all $m$ we tried.
Specifically, $\NENE = 2 \NITER+1$ in the conjugate gradient method,
because at least one line search was forced
to maintain the conjugacy of the search directions.
In contrast,
$ \NENE = \NITER+1$ in the quasi-Newton method, which means
that no line search was required in step \ref{QNLM2}.

The algorithm presented in $\S$ \ref{ALG} has a number of
options which are not uniquely determined.
Therefore, we examine some of them here,
as shown in Table \ref{TAB2}.
(a) is the reference run performed with $m=7$,
taken from Table \ref{TAB1}.
(b)-(d) were performed under the same conditions as (a)
except for the following points: 
(b) A line search with a parabolic fit was forced in each cycle
to see if the convergence rate is improved.
However, $\NENE$ was almost doubled without any reduction of $\NITER$.
Therefore, it is not justified to perform a line search
in the quasi-Newton method,
which is consistent with previous findings \cite{NOCE}.
(c) The Hessian was discarded at the end of each ionic step.
Since the convergence rate deteriorates significantly,
the inheritance of the Hessian seems to be profitable. 
(d) We tried $ \sigma_k=|\yvec_k|^2 / \yvec_k^T \svec_k $,
which gave good results in Refs. [\citen{NOCE,GILE}]. 
However, this choice requires more iterations on average,
presumably because $\sigma_k$ varies too rapidly with $k$. 
The norm of the gradient also decays
less smoothly in this case.

So far the previous search directions have been uncompressed,
i.e. stored as 64-bit double-precision arrays.
The effect of compression is examined here
in a series of runs for $m=3$ and $7$,
with different $\NBIT$.
As shown in Table \ref{TAB3}, the performance
is maintained after compression by a factor of 8-16,
especially for $m=7$.
Moreover, no instability occurred up to $\NBIT=3$.
We also show the distribution of the search direction
after compression with $\NBIT=8$
in Fig. \ref{PDOS}.
The distribution function $d(x)$ is defined as the
number of elements of $\PINT$ such that
$\PINT (i) = x$, where $ 1 \le i \le {\cal N}$ and
$ 0 \le x \le 2^8-2 = 254$.
Therefore, $\sum_x d(x) = {\cal N}$, and
there are two singularities at $x=0$ and 254.
The width of the distribution is approximately equal to $\IMAX$,
which indicates that our choice of the scale factor
(eq.(\ref{SCLF})) is appropriate. 

Finally, in order to examine the generality of
our implementation,
part of the runs were repeated for
an isolated cytosine molecule (C$_4$H$_5$N$_3$O)
in a cubic supercell of (16 a.u.)$^3$, with a timestep of 40 a.u.
($\sim$ 1 fs).
The average cutoff energy was 39 Ry, which
corresponds to $\NB=21, \NG= 8 \times 16^3 = 32,768$, and
${\cal N} \sim 700,000 $.
The results shown in Table \ref{TAB4} suggest
that the performance of the quasi-Newton method in this system
is somewhat more robust against compression,
but is qualitatively similar to the previous results in other respects.

\section{Summary}
We have shown in this article that
the quasi-Newton method using the BFGS formula
is the method of choice for
large-scale electronic-structure calculations,
if combined with efficient memory management.
The advantages of the quasi-Newton method over the conjugate
gradient method are summarized as follows:
(i) The Hessian of the previous ionic step can
be taken over to accelerate the convergence.
(ii) Practically no line search
is required, which reduces the cost of
each step significantly. 

Although there is room for fine-tuning the algorithm and
more extensive tests are necessary,
the quasi-Newton method will
provide significant speedups of the first-principles codes,
together with other techniques like
the enhanced extrapolation scheme \cite{EES} and
the constrained molecular dynamics \cite{RAT,JPSJ2}.

\section*{Acknowledgements}
The author would like to thank Dr. K.~Terakura for
helpful discussions.
The numerical calculations were performed on Hitachi SR-8000 at the
Tsukuba Advanced Computing Center.

\newpage

\begin{table}
\caption{Notation for the full/reduced vectors. Note that
$\vvec$ and $\uvec$ denote the previous and
the current gradients, respectively.}
\label{TAB0}
\begin{tabular}{lcc}
\hspace*{2cm} & \hspace{1.5cm} & \hspace{1.5cm} \\
\hline
         & full     & reduced \\
\hline
gradient         & $\gvec$    & $\vvec, \uvec$ \\
search direction & $\pvec$    & $\qvec$ \\
update vectors   & $\Delta \wfvec$  & $\svec$ \\
                 & $\Delta \gvec$   & $\yvec$ \\
wavefunction     & \wfvec     &  - \\
\hline
\end{tabular}
\end{table}

\begin{table}
\caption{
The performance of the conjugate gradient method and the
quasi-Newton method is compared in the molecular-dynamics simulations
of bulk diamond. $\NITER$ and $\NENE$ denote the number of iterations
and total energy evaluations averaged over 50 ionic steps,
respectively. }
\label{TAB1}
\begin{tabular}{rrr}
\hspace*{2.0cm} & \hspace{1.5cm} & \hspace{1.5cm} \\
\hline
method     & $\NITER$ & $\NENE$ \\
\hline
Conjugate gradient  & 14.3  & 29.6 \\
\hline
BFGS, \,\,\, $m$ =  2 & 14.9  & 15.9 \\
                    3 & 13.8  & 14.8 \\
                    4 & 13.8  & 14.8 \\
                    5 & 12.9  & 13.9 \\
                    6 & 12.4  & 13.4 \\
                    7 & 12.1  & 13.1 \\
                    8 & 11.9  & 12.9 \\
                    9 & 11.7  & 12.7 \\
                   10 & 11.6  & 12.6 \\    
                   20 & 12.4  & 13.4 \\
\hline
\end{tabular}
\end{table}

\begin{table}
\caption{
A number of variants are compared for
the BFGS with $m = 7$. (a) Reference run from Table \ref{TAB1}.
(b) A line search with a parabolic fit was forced in each cycle.
(c) The Hessian was discarded at the end of each ionic step.
(d) $ \sigma_k = |\yvec_k|^2 / \yvec_k^T \svec_k $
was used as the curvature for the new direction. }
\label{TAB2}
\begin{tabular}{ccc}
\hspace*{1.5cm} & \hspace{2cm} & \hspace{1.5cm} \\
\hline
method     & $\NITER$ & $\NENE$ \\
\hline
(a)  & 12.1  & 13.1 \\
(b)  & 12.2  & 25.4 \\
(c)  & 15.1  & 16.1 \\
(d)  & 14.3  & 15.4 \\
\hline
\end{tabular}
\end{table}

\begin{table}
\caption{
The effect of compression is compared for the BFGS with $m=3$ and 7.
}
\label{TAB3}
\begin{tabular}{rrrr}
\hspace*{1cm} & \hspace{1.2cm} & \hspace{1.5cm} & \hspace {1.5cm} \\
\hline
$m$   & $\NBIT$ & $\NITER$ & $\NENE$ \\
\hline
 3    & 64 & 13.8  & 14.8 \\
      &  8 & 14.1  & 15.1 \\
      &  4 & 14.3  & 15.3 \\
      &  3 & 15.4  & 16.4 \\
 7    & 64 & 12.1  & 13.1 \\
      &  8 & 12.2  & 13.2 \\
      &  4 & 12.2  & 13.2 \\
      &  3 & 13.3  & 14.3 \\
\hline
\end{tabular}
\end{table}

\begin{table}
\caption{
The results of selected runs from Table \ref{TAB1}-\ref{TAB3},
repeated for an isolated cytosine molecule (C$_4$H$_5$N$_3$O).}
\label{TAB4}
\begin{tabular}{rrrr}
\hspace*{2.0cm} & \hspace{1.2cm} & \hspace{1.5cm} & \hspace {1.5cm} \\
\hline
method  & $\NBIT$ & $\NITER$ & $\NENE$ \\
\hline
Conjugate gradient  & -- & 13.1  & 27.1 \\
BFGS, \,\,\, $m=3$  & 64 & 13.0  & 14.0 \\
                    &  8 & 13.0  & 14.0 \\
                    &  4 & 13.0  & 14.0 \\
                    &  3 & 13.3  & 14.3 \\
BFGS, \,\,\, $m=7$  & 64 & 10.7  & 11.7 \\
\hline
\end{tabular}
\end{table}

\begin{figure}
   \caption{The distribution function $d(x)$ of the compressed search
   direction $\PINT$ for $\NBIT = 8$.
   }
   \label{PDOS}
\end{figure}


\begin{thebibliography}{99}
  \bibitem{HK} P.~Hohenberg and W.~Kohn: Phys.~Rev. {\bf 136} (1964) B864.

  \bibitem{KS} W.~Kohn and L.~J.~Sham: Phys.~Rev. {\bf 140} (1965) A1133.

  \bibitem{CP} R.~Car and M.~Parrinello:
  Phys.~Rev.~Lett. {\bf 55} (1985) 2471.

  \bibitem{PAY} M.~C.~Payne, M.~P.~Teter, D.~C.~Allan, T.~A.~Arias
  and J.~D.~Joannopoulos: Rev.~Mod.~Phys. {\bf 64} (1992) 1045.

  \bibitem{TREV} K.~Terakura: {\it Computational Physics
  as a New Frontier in Condensed Matter Research}, ed. H.~Takayama,
  M.~Tsukada, H.~Shiba, F.~Yonezawa, M.~Imada
  and Y.~Okabe (The Physical Society of Japan, Tokyo, 1995).

  \bibitem{MARX} D.~Marx and J.~Hutter: {\it Modern Methods and
  Algorithms of Quantum Chemistry}, ed. J.~Grotendorst (NIC,
  Forschungszentrum Julich, 2000), available at
  http:// www.theochem.ruhr-uni-bochum.de/
  research/ marx/ index.en.html.

  \bibitem{CTS} J.~R.~Chelikowsky, N.~Troullier, K.~Wu and Y.~Saad:
  Phys.~Rev.~B {\bf 50} (1994) 11355.

  \bibitem{BER} E.~L.~Briggs, D.~J.~Sullivan and J.~Bernholc:
  Phys.~Rev.~B {\bf 54} (1996) 14362.

  \bibitem{RREV} T.~L.~Beck: Rev.~Mod.~Phys. {\bf 72} (2000) 1041.

  \bibitem{WHT} S.~R.~White, J.~W.~Wilkins and M.~P.~Teter:
  Phys.~Rev.~B {\bf 39} (1989) 5819.

  \bibitem{PRB2} E.~Tsuchida and M.~Tsukada:
  Phys.~Rev.~B {\bf 54} (1996) 7602.

  \bibitem{JPSJ} E.~Tsuchida and M.~Tsukada:
  J.~Phys.~Soc.~Jpn. {\bf 67} (1998) 3844,
  available at http:// wwwsoc.nii.ac.jp/ jps/ jpsj/.

  \bibitem{PKFS} J.~E.~Pask, B.~M.~Klein, C.~Y.~Fong and P.~A.~Sterne:
  Phys.~Rev.~B {\bf 59} (1999) 12352.

  \bibitem{RCP} W.~H.~Press, S.~A.~Teukolsky, W.~T.~Vetterling
  and B.~P.~Flannery: {\it Numerical Recipes in Fortran}
  (Cambridge University Press, Cambridge, 1992).

  \bibitem{GLL} M.~J.~Gillan:
  J.~Phys.:~Condens.~Matter {\bf 1} (1989) 689.

  \bibitem{TPA} M.~P.~Teter, M.~C.~Payne and D.~C.~Allan:
  Phys.~Rev.~B {\bf 40} (1989) 12255.

  \bibitem{SCP} I.~Stich, R.~Car, M.~Parrinello and S.~Baroni:
  Phys.~Rev.~B {\bf 39} (1989) 4997.

  \bibitem{BKL} D.~M.~Bylander, L.~Kleinman and S.~Lee:
  Phys.~Rev.~B {\bf 42} (1990) 1394.

  \bibitem{PLY} P.~Pulay: Chem.~Phys.~Lett. {\bf 73} (1980) 393.

  \bibitem{WOZU} D.~M.~Wood and A.~Zunger:
  J.~Phys.~A {\bf 18} (1985) 1343.

  \bibitem{MACO} J.~L.~Martins and M.~L.~Cohen,
  Phys.~Rev.~B {\bf 37} (1988) 6134.

  \bibitem{HLP} J.~Hutter, H.~P.~L\"uthi and M.~Parrinello:
  Comp.~Mat.~Sci. {\bf 2} (1994) 244.

  \bibitem{KRFU} G.~Kresse and J.~Furthm\"uller:
  Comp.~Mat.~Sci. {\bf 6} (1996) 15.

  \bibitem{SREV} T.~Schlick: Rev.~Comp.~Chem. {\bf 3} (1992) 1.

  \bibitem{FREV} R.~Fletcher: Report NA/149, Department of
   mathematics and computer science, University of Dundee, 1993,
   availble at
   http:// citeseer.nj.nec.com/ fletcher93overview.html.

  \bibitem{NREV} J.~Nocedal: {\it The State of the Art in
  Numerical Analysis}, ed. I.~S.~Duff and G.~A.~Watson
  (Oxford University Press, 1998), available at
  http:// www.ece.nwu.edu/\~{}nocedal/ recent\_{}pub.html.

  \bibitem{HEPO} M.~Head-Gordon and J.~A.~Pople:
  J.~Phys.~Chem. {\bf 92} (1988) 3063.

  \bibitem{FIAL} T.~H.~Fischer and J.~Alml\"of:
  J.~Phys.~Chem. {\bf 96} (1992) 9768.

  \bibitem{HSZ} R.~A.~Hyman, M.~D.~Stiles and A.~Zangwill:
  Phys.~Rev.~B {\bf 62} (2000) 15521.

  \bibitem{NOCE} D.~C.~Liu and J.~Nocedal:
  Math.~Prog. {\bf 45} (1989) 503.

  \bibitem{SIEG} D.~Siegel: Report DAMTP 1992/NA12,
  University of Cambridge.

  \bibitem{GILE} P.~E.~Gill and M.~W.~Leonard:
  Report NA 97-1, Department of Mathematics, Santa Clara University,
  available at http:// www.math.ucla.edu/
  \~{}mwl/ vita/ vita.html.

  \bibitem{APJ1} T.~A.~Arias, M.~C.~Payne and J.~D.~Joannopoulos:
  Phys.~Rev.~Lett. {\bf 69} (1992) 1077.

  \bibitem{MGC} F.~Mauri, G.~Galli and R.~Car:
  Phys.~Rev.~B {\bf 47} (1993) 9973.

  \bibitem{KV} R.~D.~King-Smith and D.~Vanderbilt:
  Phys.~Rev.~B {\bf 49} (1994) 5828.

  \bibitem{WM} See, {\it e.g.}, R.~M.~Wentzcovitch and J.~L.~Martins:
  Solid State Commun. {\bf 78} (1991) 831.

  \bibitem{KB} L.~Kleinman and D.~M.~Bylander:
  Phys.~Rev.~Lett. {\bf 48} (1982) 1425.

  \bibitem{GTH} S.~Goedecker, M.~Teter and J.~Hutter:
  Phys.~Rev.~B {\bf 54} (1996) 1703.

  \bibitem{EES} E.~Tsuchida and K.~Terakura:
  J.~Phys.~Soc.~Jpn. {\bf 71} (2002), to appear.

  \bibitem{LIQ} M.~P.~Allen and D.~J.~Tildesley:
  {\it Computer Simulation of Liquids} (Oxford Science Publications,
  Clarendon, Oxford, 1987).

  \bibitem{RAT} H.~C.~Andersen: J.~Comp.~Phys. {\bf 52} (1983) 24.

  \bibitem{JPSJ2} E.~Tsuchida and K.~Terakura:
  J.~Phys.~Soc.~Jpn. {\bf 70} (2001) 924.

\end{thebibliography}
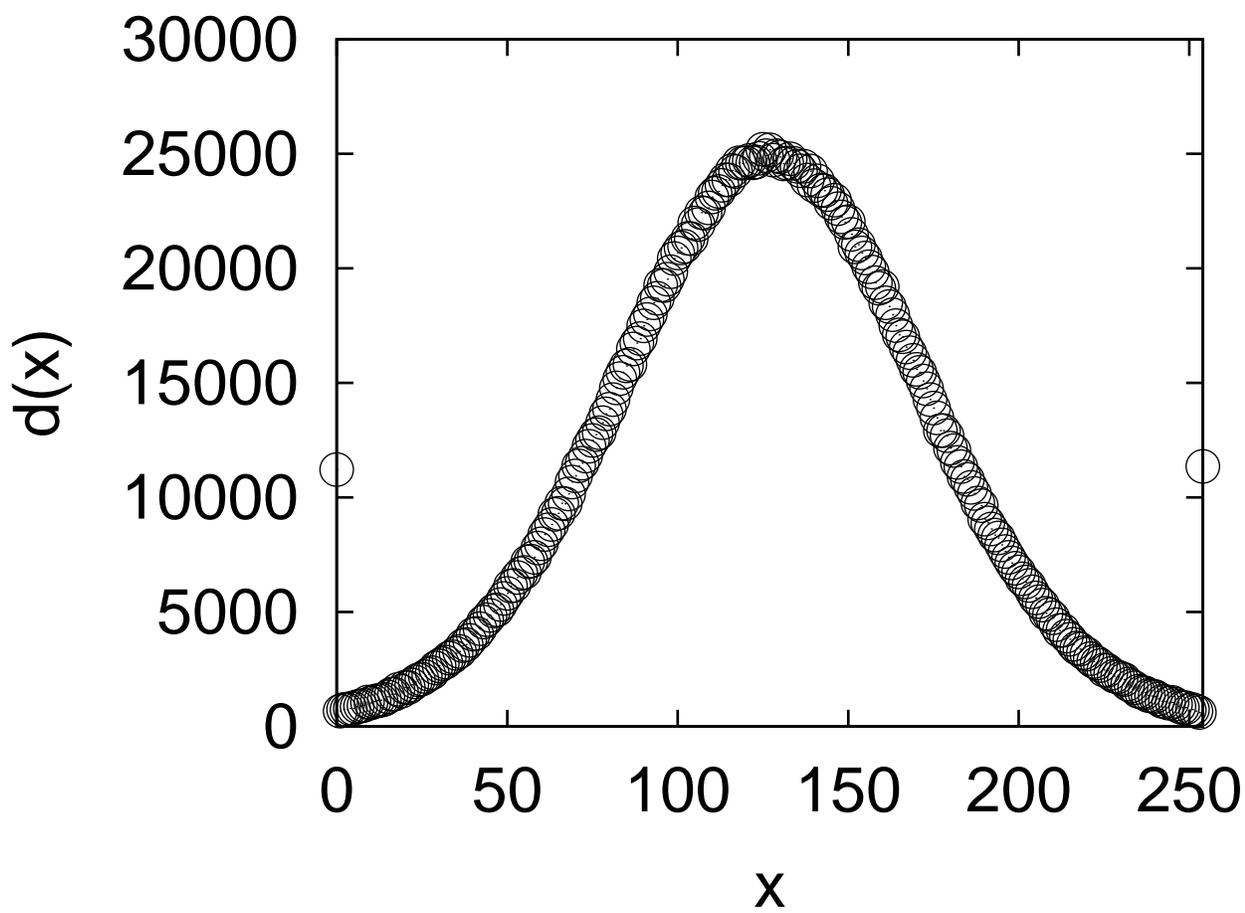
\end{document}